\documentclass[singlespacing]{elsart}

\usepackage{graphicx}

\usepackage{amssymb}
\journal{}
\begin{document}

\begin{frontmatter}

\title{Solvent viscosity dependence for enzymatic reactions}

\author{A.E. Sitnitsky},
\ead{sitnitsky@mail.knc.ru}

\address{Institute of Biochemistry and Biophysics, P.O.B. 30, Kazan
420111, Russia. e-mail: sitnitsky@mail.knc.ru }

\begin{abstract}
A mechanism for relationship of solvent viscosity with reaction rate constant
at enzyme action is suggested. It is based on
fluctuations of electric field in enzyme active site produced by
 thermally equilibrium rocking (cranckshaft motion) of the rigid
plane (in which the dipole moment $\approx 3.6\ D$ lies) of a
favourably located and oriented peptide group
(or may be a few of them). Thus the rocking of the plane
leads to fluctuations of the electric field of the dipole moment.
 These fluctuations can
interact with the reaction coordinate because the latter in its turn has
transition dipole moment due to separation of charges at movement of the
reacting system along it. The rocking of the plane of the peptide group
is sensitive to the microviscosity of its environment in protein interior
 and the latter is a function of the solvent viscosity. Thus we obtain an
additional factor of
interrelationship for these characteristics with the reaction rate constant.
We argue that due to the properties of the cranckshaft motion
the frequency spectrum of the electric field fluctuations has
a sharp resonance peak at some frequency and the corresponding Fourier mode
can be approximated as oscillations. We employ a known result from the
theory of thermally activated escape with periodic driving to obtain the
reaction rate constant and argue that it yields reliable description of the preexponent
where the dependence on solvent viscosity manifests itself.
The suggested mechanism is shown to grasp the main feature of this
dependence known from the experiment
and satisfactorily yields the upper limit of the fractional
index of a power in it.
\end{abstract}

\end{frontmatter}

\section{Introduction}
The functional dependence of the rate
limiting stage
$k_{cat}$ for enzymatic and protein (ligand binding/rebinding)
reactions on solvent viscosity $\eta$
 of the type
\begin{equation}
\label{eq1} k_{cat}\propto \frac{1}{\eta^{\beta}}
\end{equation}
where $ 0 < \beta < 1$ (usually $\beta \approx 0.4 \div 0.8$)
 has been known for a long time  \cite{Gav79}, \cite{Bee80},
 \cite{Fra88},  \cite{Dem89}, \cite{Fra90}, \cite{Ng91},
\cite{Ng911}, \cite{Yed95}, \cite{Dos95}, \cite{Kle98}, \cite{Fra99}.
More detailed studies revealed that in fact the fractional
index of a power $\beta$
is a function of cosolvent molecular weight $M$ (i.e., the mass of
a cosolvent molecule expessed in atomic units and measured in Daltons) \cite{Yed95}
\begin{equation}
\label{eq2} \beta=\beta(M)
\end{equation}
If one varies the solvent viscosity by large cosolvent molecules with high
molecular weight that do not penetrate into enzyme then one obtains that the
fractional exponent $\beta \rightarrow 0$, i.e., the reaction rate constant does
not depend on solvent viscosity. With the decrease of cosolvent molecular
weight the fractional exponent $\beta$ increases. In the limit of
hypothetical "ideal" cosolvent with infinitely small molecular weight
(cosolvent molecules freely penetrate into enzyme and are distributed
there homogeneously) it
tends to a limit value $\beta_{max}=\lim_{M\to 0}{\beta(M)}\approx 0.79$.
The latter is neither experimental value nor a calculated one.
It is an extrapolated number (see \cite{Yed95} for details).

The functional dependence (\ref{eq1}) also takes place for folding of proteins
(see \cite{Pab04}, \cite{Fra06} and refs therein). In our opinion enzyme
catalysis and folding are quite different phenomena proceeding on different
timescales (an enzyme turnover is typically $10^{-4} \div 10^{-3}\ s$ while
"the time of folding varies from microseconds to hours" \cite{Fra06}).
An enzymatic or protein reaction typically has a distinct rate limiting stage and can
be perceived as an elementary step. Folding is a complicated process that
involves huge number of elementary steps of commensurable importance.
That is
why we suppose that enzyme catalysis and folding involve different origin of
 the dependence
(\ref{eq1}). In the present paper we deal only with enzymatic and protein
reactions (i.e., those of bond breaking/bond making in protein interior)
 and do not touch upon folding. For an unprejudiced observer folding seems to be
 an overworked
issue in the literature while physical aspects of enzyme action still remain in a
deep shadow of their chemical counterparts for this phenomenon. Overwhwelming majority
of researches from physical community perceive enzyme catalysis as some "chemistry"
or "biology". That is why the aim to attract their attention to it as to a
physical problem initiated by the the collection \cite{Flu85} and continued by
the review article \cite{Fra99} seems to remain as urgent as it was in the
 previous century.

The attempts to explain the functional dependence (\ref{eq1}) can be
roughly divided into "phenomenological" and "theoretical". The former
suggest that the fractional exponent $\beta$ is the degree with which solvent
viscosity is coupled with (frequency dependent friction) \cite{Dos83}
 or penetrates into  (position dependent friction) \cite{Gav80},
  \cite{Bar95} the protein interior. The latter try to
 derive it from the first principles
\cite{Gro80}, \cite{Zwa92}. However Zwanzig model yields too small value for
the fractional exponent $\beta=0.5$ \cite{Zwa92}. Grote-Hynes theory \cite{Gro80}
gives that the rate dependence on solvent viscosity should be weaker than
that predicted by Kramers' one (the latter yields $k\propto 1/\eta$
in the high friction limit \cite{Han90}). However no explicit derivation of
expression (\ref{eq1}) from the Grote-Hynes theory has been achieved.
As the authors of \cite{Yed95} conclude "there seems to be no general
agreement yet about the origin of the fractional $\beta$ value in Eq.1".
The authors of \cite{Kle98} draw to a similar conclusion.
In our opinion little has changed in this issue (as applied to enzyme
catalysis only
because there is certain progress in understanding of viscosity dependence for
folding \cite{Fra06}) since the date of
the cited papers. The aim of the present paper is to provide theoretical
interpretation of the functional dependence (\ref{eq1}) and to
"explain" the limit value $\beta_{max}\approx 0.79$.

There seems to be a consensus among researhes in understanding that the
dependence of an enzymatic reaction rate constant on solvent viscosity is mediated by
internal protein dynamics. This undersanding goes back to the so called transient strain
model. The latter is based on the idea of overcoming the energy
barrier of an enzymic reaction by structural fluctuations whose frequency
is inversely proportional to the viscosity of the medium \cite{Gav78},
\cite{Gav86}, \cite{Gav94}.
 That is why any theory of the phenomenon should be
a part of the mainstream of modern enzymology to study the
role of dynamical contribution into enzyme catalysis (see the materials of
 a recent conference in
the subject issue of Phil. Trans. R. Soc. B (2006) 361). There are
different sonorous names for such dynamical mechanism:
 "rate promoting vibration"
 (RPV) \cite{Ant06},  the "protein promoting modes" \cite{Agr05},
\cite{Agr06}, etc.
In the present paper the name RPV is used as the most appropriate one
 for the concept under consideration that some conformational motion of
 vibrational character in
protein is coupled somehow to the reaction coordinate.
However it should be stressed that the author of the present paper input in
 this name absolutely different
meaning than the authors of \cite{Ant06}, \cite{Agr05}, \cite{Agr06} and other
papers within the framework of this concept. We invoke to the
idea that a dynamically unusual
 electric field in enzyme active site may play a key role for catalysis.
This idea was put forward by Fr\"ohlich in his concept of coherent
vibrations of protein giant dipole moment \cite{Fro75}, \cite{Fro86},
Gavish and Werber in the hypothesis of charge fluctuations \cite{Gav79} and
Warshel in his concept of electrostatic fluctuations \cite{War78},
\cite{War84}
(see also \cite{Ols06}, \cite{Ols006},
 \cite{War06} and refs. therein).
In the present paper the name RPV means the following: {\sl
a Fourier mode of
the fluctuating electric field in the enzyme
 active site generated by protein dynamics}
 \cite{Sit95}, \cite{Sit06}.

 Warshel and coauthors \cite{Ols06}, \cite{Ols006},
 \cite{War06} argue the following statements.
1. A dynamical mechanism can contribute significantly
 into enzyme catalytic efficiency only if it leads to nonequilibrium
 (non-Boltzmann) distributions for the reaction coordinate produced
 by coherent oscillations in protein dynamics.
 In the opposite case of thermal equilibrium  dynamical
effects can lead to nothing more than some modest corrections in the
 preexponent.
2.  Equilibrium protein dynamics can not lead to coherent
oscillations coupled to the reaction coordinate.
The item 2. from this list is doubtless. However the item 1. in our
opinion is not so and an efficient dynamical mechanism can stem from
thermally equilibrium fluctuations. Moreover even if the item 1. is
true (i.e., a dynamical mechanism
does not contribute into the catalytic efficiency)
the corrections in the
preexponent can be crucial for the dependence of the enzymatic reaction rate
constant on solvent viscosity because the latter manifests itself namely in the
preexponent. We discuss a possibility that protein dynamics
produces specific fluctuational influence on the reaction coordinate. This influence on
the one hand is of thermally equilibrium origin and on the other hand is
additional to those available for reactions in solution
(i.e., the latter have no analogous counterpart in the thermal noise
 spectrum). Of course this
 fluctuational influence  can not be coherent oscillations.
However in our opinion coherent oscillations are not crucial to be
 the origin of
the dynamical mechanism. Fluctuations of thermally equilibrium nature can
play this role as well. Regretfully as will be argued below it seems
 rather difficult to treat such fluctuations in their natural form.
That is why in the present paper we invoke to the fact that
some Fourier mode from their frequency spectrum mimic coherent
oscillations so much that can be considered in the first
approximation as a steady harmonic vibration. We stress once more
that it is merely
a methodical trick to reduce the problem to elaborate theoretical technique
rather than an indispensable assumption for the present approach.
 The question "how much can
such vibration
contribute into the reaction rate
enhancement ?" is not the matter of the present paper and is
touched upon rather briefly here. The aim of the paper is to show that this vibration
(that is the RPV in our approach) enables us to
interprete
experimental data on solvent viscosity depenedence for enzymatic reactions.

The paper is organized as follows. In Sec. 2 the relevant protein
dynamics as the origin of
the electric field fluctuations is introduced.
In Sec. 3 the interaction of these fluctuations with the
reaction coordinate is disscused.
In Sec.4 the equations of motion are obtained and
the reasons why the electric field fluctuations
 can be conceived as oscillations are argued.
In  Sec. 5 the influence of
these oscillations on the reaction rate is considered. In Sec. 6
numerical estimates are presented. In Sec. 7 the solvent
viscosity dependence for the
reaction is obtained. In Sec.8 the results are discussed and the
conclusions are summarised. In Appenix A some technical details are presented.
In Appendix B some additional material is
presented.

\section{Origin of the RPV}
A crucial question for the mainstream of modern enzymology to investigate
the role of dynamical effects at enzyme catalysis is the following: how does
the RPV that is typically on the picosecond time scale affect
the catalytic act that is typically on the millisecond time scale (an enzyme
turnover is usually $\approx 10^{-3}\div 10^{-4} s$)
(see \cite{Ant06} and refs. therein)? In our opinion
 the frequency of the RPV as itself is not an appropriate
characteristic for this issue. As a matter of fact it is not of much
significance whether the RPV is on the picosecond time scale or, e.g., on
the nanosecond time scale. The relevant and the most important characteristic
 is the life time of vibrational motion in protein dynamics that produces
 the RPV. Namely the problem of survival of vibrational excitations on
the time scale of enzyme turnover plagues many of the
speculations about dynamical contribution into enzyme action and is a point of
application for criticism by Warshel and coauthors \cite{Ols06}, \cite{Ols006},
 \cite{War06}.
It is argued below that this problem does not
arise in the present approach.

The most natural candidate for the source of
electric field is a peptide group of protein backbone.
The latter is known to have a rather large constant
dipole moment $\bar p \approx 3.6\ D$ that lies in its plane \cite{Can80}.
The dipole moment produces an electric field.
Thermal fluctuations (rocking) of the rigid plane of the peptide group
relative to its mean averaged position in the protein backbone
lead to variation in time of this electric field, i.e., to the electric field
fluctuations. As the latter are  produced by thermally
 equilibrium fluctuations they exist on the whole
duration time of the catalytic act. That is why there is no problem to
match the electric field
 fluctuations with the process of catalysis. We are interested in
the amplitude and spectral properties of these fluctuations in the enzyme
active site at the place of the reaction coordinate.

The rocking of the rigid plane of the peptide group (the so called
"crankshaft-like" motion) due to degrees of freedom of torsional (dihedral)
 angles $\varphi_i$ and $\psi_{i-1}$ (see Fig.1) was
proposed on theoretical grounds from normal-mode analysis \cite{Go76}.
It is supported by numerous NMR experiments and molecular
dynamics simulations of protein backbone  \cite{McC77},
 \cite{Lev79}, \cite{Gun82}, \cite{Lev83}, \cite{Del89}, \cite{Cha92}, \cite{Pal92},
\cite{Bru95}, \cite{Fad95}, \cite{Buc99},
\cite{Der99}, \cite{Ulm03}, \cite{Clo04}, \cite{Fit07}.
The crankshaft-like motion is comprehended now as a dominant type of motion
for the protein backbone that
"involves only a localized oscillation of the plane of the peptide group"
 \cite{Fit07}. The essence of this motion is the so called anticorrelated
motion of the torsional angles $\varphi_i$ and $\psi_{i-1}$
manifested itself in the requirement (see Fig.2)
\begin{equation}
\label{eq3}
\phi/2\equiv \Delta \varphi_i = \Delta \psi_{i-1}
\end{equation}
In this case the plane of the
peptide group rocks as a whole around some axis $\sigma$ that goes through the
center of masses of the peptide group parallel to the bonds
$C_{\alpha}^{i-1}-C^i$ and $N^i-C_{\alpha}^{i}$ (see Fig.1).
The moment of
inertia of the peptide group relative to the axis $\sigma$
is known to be $I \approx 7.34 \cdot 10^{-39}
g \cdot cm^{2}$ \cite{Flu94}. However this value is given in \cite{Flu94}
without a reference or calculation. That is why in Appendix A a brief
estimate corroborating this value is given.
Molecular dynamics simulations and NMR experimental data testify  that the character of
the correlation function for the crankshaft motion is  decaying
oscillations \cite{Gun82}, \cite{Fit07}. However they provide characteristics
for them in a very wide range from the subpicosecond
and picosecond time scale \cite{Fit07} to  slower motions on a much larger
time scale from tens of picoseconds to 100 ps and more \cite{Clo04}.
This is presumably a particular manifistation for the cranckshaft motion of
a general principle of hierarchical structure for a conformational
potential in protein dynamics \cite{Fra88}, \cite{Fra90}, \cite{Fra99} when a group of
local minima forms a smooth local minimum and so on. In fact  knowing
the actual values of the frequency for the oscillations and the characteristic time of
their decay for the functionally important crankshaft motion is not
indespesible for the purposes of the present paper. However in our opinion
it is reasonable to assume that the frequency of oscillations of the plane
of the peptide group as a whole for such motion should be at least
 order of magnitude
less than those for high frequency in - plane motions such as,
 e.g., Amide I ($\sim 1600\ cm^{-1}$). The
choice of the frequency in the
$\omega_0 \sim 100\ cm^{-1}\approx 10^{13}\ s^{-1}$ range enables us to match it
with the amplitudes of rocking of order of several degrees (see (\ref{eq12})
 below) in accordance with experimental data \cite{Buc99}, \cite{Ulm03}.

It is natural to describe the cranckshaft motion by a Langevin equation.
Such equations are frequently used in protein dynamics
\cite{Ros79}, \cite{Dos06}, \cite{Rub04}.
The equation of the cranckshaft motion for the rigid plane of the peptide
 group in the above conditions is
\begin{equation}
\label{eq4} I\frac{d^2 \phi(t)}{dt^2}+\gamma \frac{d \phi(t)}{dt}+
I\omega_0^2\phi(t)=\xi(t)
\end{equation}
where $\gamma$ is the friction coefficient and $\xi(t)$ is the random torque
with zero mean $<\xi(t)=0>$ and correlation function
\begin{equation}
\label{eq5} <\xi(0)\xi(t)>=2k_BT\gamma \delta(t)
\end{equation}
 $k_B$ is the Boltzman constant, $T$ is the temperature.

We consider the hydrodynamic friction. This "macroscopic"
notion is known to work surprisingly well at the molecular level
(see \cite{Met77} for thorough discussion).
We model the peptide group by an oblate
ellipsoid with halfaxes $a$, $b$ and $c$ (where $a \sim c $
 and $a, c >> b$
 that reflects the flat character of the peptide group) Fig. 3.
 For the rotation of the ellipsoid around $x$-axis
 (that is in our case actually
 the axis $\sigma$ for the peptide group introduced above)
the friction coefficient is given by a formula \cite{Klu97}
\begin{equation}
\label{eq6} (\gamma)_x=8\pi \eta \ abc\left[\frac{12 r}{a}+
\frac{(bc)^{1/4}}{a^{1/2}\left(1+
\frac{4r}{a^{1/2}(bc)^{1/4}}\right)^3}\right]^{-1}
\end{equation}
where $r=2(abc)^{1/3}$ and $\eta$ is the viscosity of the
 ellipsoid environment. The required behavior is obtained if we have
 the condition of the underdamped motion
\begin{equation}
\label{eq7} \frac{\gamma}{I\omega_0} << 1
\end{equation}
Let us meke a numerical estimate of this requirement. Taking the linear size
of the peptide group (distance for the atom $O$ to the atom $H$) to be
$L\approx 2.5\ \AA$ we have $c\approx L/2$, $a\approx c/2$ and $b\approx
a/3$. From (\ref{eq6}) we obtain that at $\eta\approx 1\ cP$
(that is the viscosity of water at room temperature)
$\gamma \approx 2\cdot 10^{-27}\ CGS$. Then at typical value
 $\omega_0 \approx 10^{13} s^{-1}$ and the value $I \approx 7.34 \cdot 10^{-39}
g \cdot cm^{2}$ we have $ \gamma /(I\omega_0)\approx 0.01$, i.e., the
requirement (\ref{eq7}) is by far satisfied. Moreover it well holds even at
the increase of $\eta$ by an order of magnitude.

At the requirement (\ref{eq7}) we obtain (see e.g. \cite{Chir80}, \cite{Rub04})
\begin{equation}
\label{eq8} \alpha(t)\equiv <\phi(0)\phi(t)>\approx
 \frac{k_BT}{I\omega_0^2}exp\left(-\frac{\gamma \mid t \mid}{2I}\right)
cos\left(\omega_0t\right)
\end{equation}
We denote
\begin{equation}
\label{eq9} \mu=\frac{\gamma}{\sqrt 2 I\omega_0}
\end{equation}
Then the Fourier spectrum for the correlation function $\alpha(t)$ is
\begin{equation}
\label{eq10} \tilde \alpha(\omega)=\frac{k_BT}{I\omega_0^2}
\frac{\mu}{\pi}\frac{1+\mu^2+\left(\omega/\omega_0\right)^2}
{\mu^4+2\mu^2\left[1+\left(\omega/\omega_0\right)^2\right]+
\left[1-\left(\omega/\omega_0\right)^2\right]^2}
\end{equation}
At our requirement (\ref{eq7}) it has a sharp resonance peak
at the frequency $\omega_0$.
For the mean squared amplitude ($msa$) we have
\begin{equation}
\label{eq11} msa=\frac{k_BT}{I\omega_0^2}
\end{equation}
The latter means that the amplitude of the cranckshaft motion at room
temperature and, e.g., typical value $\omega_0 \approx 10^{13} s^{-1}$ is
\begin{equation}
\label{eq12} \phi_{max}=\sqrt{\frac{k_BT}{I\omega_0^2}}
\approx 0.1\approx 6^{\circ}
\end{equation}
that is $\Delta \varphi_i = \Delta \psi_{i-1}\approx 3^{\circ}$.
This value sheds light on the origin of essentially vibrational character of
the peptide group motion manifested itself in (\ref{eq8}) and (\ref{eq10}).
 At such angles of
rotational deviation of the peptide group from its mean averaged position
the linear displacements of the atoms are
$\sim c \cdot \phi_{max}\approx 0.1\ \AA$. The latter value is much less than
both the size
of a solvent molecule and interatomic distances to
neighbor fragments of protein structure. That is why the environment
exerts rather weak friction for such type of the peptide group motion that is
reflected in the requirement (\ref{eq7}). Thus we conclude that the
thermally equilibrium cranckshaft motion of the peptide group is of
essentially vibrational character even in such
condensed medium as protein interior.

\section{Essence of the RPV}
We consider the plane of the peptide group that undergoes thermally
equilibrium rocking around its mean averaged position. The angle
$\phi(t)$ quantifies the random deviations of the plane. We choose the axis
$x$ in the direction of these deviations (see Fig.4). Now we recall that
there is the dipole moment of the peptide group with the absolute
value $ p\approx 3.6\ D $
laying in its plane. We choose the axis $z$ in the direction of the dipole
moment at the mean averaged position of the peptide group. Thus the dipole
moment is a vector in our frame
\begin{equation}
\label{eq13} \bar p(t)=p\ sin\phi(t)\ \bar e_x+p\ cos\phi(t)\ \bar e_z
\end{equation}
Taking into account that $\mid \phi (t)\mid <<1$ we have
\begin{equation}
\label{eq14} \bar p(t)\approx p\ \phi(t)\ \bar e_x+p\ \bar e_z
\end{equation}
We assume a simplest geometry favorable for catalysis
(generalizations are trivial but lead to more cumbersome formulas
 of not principle character). In this geometry at a distance $R$ from
 the dipole moment along the axis $x$
\begin{equation}
\label{eq15} \bar R= R\ \bar e_x
\end{equation}
there is a reaction coordinate $q$. The latter has a transition
dipole moment $\bar d$ due to separation of charges with the value $e$ at
movement along it
\begin{equation}
\label{eq16} \bar d =e(q-q_0)\bar e_x
\end{equation}
Here $q_0$ is the position of the minimum of the potential energy surface
corresponding to initial reagent.
Then the quasistationary fluctuating electric field of the peptide group dipole
moment in the place of the reaction coordinate is
\begin{equation}
\label{eq17} \bar E(t)=\frac{3\left(\bar p(t)
\cdot \bar R \right) \bar R-R^2\bar p(t)}
{\epsilon R^5}
\approx \frac{p}{\epsilon R^3}
\left[2\phi(t)\bar e_x-\bar e_z\right]
\end{equation}
where $\epsilon$ is the dielectric constant. Denoting
\begin{equation}
\label{eq18} g=\frac{2ep}{\epsilon R^3}
\end{equation}
we obtain for the interaction of the peptide group dipole moment with the
transition dipole moment of the reaction coordinate
\begin{equation}
\label{eq19} H_{int}=-\left(\bar d \cdot \bar E(t)\right)=-g(q-q_0)\phi(t)
\end{equation}
 This interaction reveals the essence of the RPV as
fluctuating electric field affecting the reaction coordinate.

\section{Equations of motion}
We denote $m$ the effective mass of the reaction coordinate $q$ and
$V(q)$ the potential energy surface for it. We assume the most
frequently
used memory-free friction  (Ohmic damping) for both the
reaction coordinate and the peptide group cranckshaft motion. For the
reaction coordinate it can be motivated as follows.
 Taking into account the memory friction (that leads  to
the generalized Langevin equation, e.g., in the spirit of
Grote-Hynes approach) is usually necessary for fast reactions (e.g.,
those with low activation barriers). Enzymatic reactions typically have
a rather high barrier (even after its reduction by catalytically active
groups). Thus the limit of a very rapidly fluctuating force compared with
the rate of the transition over the barrier (the reaction
dynamics is very slow and the environment exerts its full frictional
influence during barrier crossing) can be safely taken. In this
case the generalized Langevin equation is well approximated by the ordinary
one. For the high - frequency cranckshaft motion of the peptide group the
approximation of Ohmic damping
can not be logicaly motivated because the characteristics of the correlation
function (inverse frequency of oscillations and decay time) are comparable
with the characterisic time of surrounding rearrangements acting as thermal
bath. Besides taking into account memory friction for peptide dynamics can
be implemented rather easily in the approximation of harmonic
conformational potential
\cite{Dos83}, \cite{Dos06}, \cite{Sit96}, \cite{Sit00},
\cite{Sit02}. However it usually leads to complications of not
principal character at room temperatures (well above the so called
"glass" transition in protein dynamics) where enzymes normally work.
In this temperature range conformational protein dynamics are satisfactorily
described by ordinary Langevin equation.

Then we have classical equations of
motion are
\begin{equation}
\label{eq20} m \ddot q+V^{\prime}(q)-g\phi+\nu \dot q =\zeta(t)
\end{equation}
\begin{equation}
\label{eq21} I \ddot \phi+I\omega^2_0 \phi-g(q-q_0)+
\gamma \dot \phi =\xi(t)
\end{equation}
where the coupling constant $g$ is given by (\ref{eq18}).
In (\ref{eq20}) $\nu$ is the friction coefficient for translational motion along the
reaction coordinate (modeled by effective Brounian particle with
 the caharacterisic linear size $l$ in the media with viscosity $\eta$
 and given by the Stokes formula $\nu=6\pi l \eta$ (see next Sec.)) and
$\zeta(t)$ is the random force characterized by
\begin{equation}
\label{eq22}<\zeta(t)>=0; \ \ \ \ \ \ \ \ \ \ \
<\zeta(0)\zeta(t)>=2 k_BT\nu \delta(t)
\end{equation}
In (\ref{eq21}) $\gamma$ is the friction coefficient for rotational fluctuations of
the peptide group given by (\ref{eq6}) and $\xi(t)$ is the random torque
characterized by
\begin{equation}
\label{eq23}<\xi(t)>=0; \ \ \ \ \ \ \ \ \ \ \
<\xi(0)\xi(t)>=2 k_BT\gamma \delta(t)
\end{equation}
Neglecting the "backward" influence of the reaction coordinate
on the peptide group motion (term $-g(q-q_0)$ in (\ref{eq21})) we
 obtain for the latter the previously considered equation
of motion (\ref{eq4}) and are left with a two noise problem
\begin{equation}
\label{eq24} m \ddot q+V^{\prime}(q)+\nu \dot q =\zeta(t)+g\phi(t)
\end{equation}
Here the internal (its intensity is related with friction coefficient
 $\nu$) white noise $\zeta(t)$ is characterized by
(\ref{eq22}) and the external (it does not create friction for the
 movement along the reaction coordinate $q$)
oscillating noise $\phi(t)$ is characterized by
\begin{equation}
\label{eq25}<\phi(t)>=0; \ \ \ \alpha(t)\equiv <\phi(0)\phi(t)>\approx
 \frac{k_BT}{I\omega_0^2}exp\left(-\frac{\gamma \mid t \mid}{2I}\right)
cos\left(\omega_0t\right)
\end{equation}
The stochastic influence $g\phi(t)$ has no counterpart for reactions in
solution. It is unique for enzymatic reactions because it is produced by
dynamics of protein structure (namely by a favourably located and oriented
 peptide group in the present model or may be by a few of them in more
realistic cases). In solution solvent molecules also posess dipole moments
and undergo thermal motion. However they do not form strictly determined
structure enabling them to
implement high-frequency motion of essentially vibrational character as in
the case of peptide groups in proteins.

The equations (\ref{eq24}), (\ref{eq22}), (\ref{eq25}) belong to a class of
the problems considered recently in \cite{Ray01},
 \cite{Ray07}. Regretfully the formulas obtained there lead to very
cumbersome manipulations in our case and this most natural way of
 formulating the problem has not led to representable results for
oscillating noise (\ref{eq25}) yet.
That is why we have to resort to simplifying assumptions. First of all we
take into account that a reaction of bond breaking or bond making requires
linear displacements of atoms of order of the bond length
 $\sim 1\ \AA$ that is comparable with the size of solvent molecules.
Hence the solvent exerts rather strong friction for the movement of the
system along the reaction coordinate. Thus we can restrict ourselves by the
high friction limit, i.e., neglect the inertial term $ m \ddot q$ in
(\ref{eq24}). The requirement for the overdamped
regime is $\nu/(m\omega_b) >> 1$ where  the friction coefficient is given by
the Stokes formula $\nu = 6\pi l\eta$ (see next Sec.) and the frequency $\omega_b$
characterizes the shape of the potential barrier at its top
$\omega_b = \sqrt {\mid V^{\prime \prime}(q_b)\mid/m}$.
At $l\approx 1\ \AA$ and $\eta\approx 1\ cP$
(that is the viscosity of water at room temperature) we have the estimate
 $\nu\approx 2\cdot 10^{-9}\ CGS$. Then at typical values
 $\omega_b \sim 10^3\ cm^{-1} \approx 10^{14}\ s^{-1}$ and $m \approx 5\ a.u.m.$
($a.u.m.\approx 1.7 \cdot 10^{-24} g$)
 we obtain $\nu/(m \omega_b) \approx 2$. Thus the requirement of overdamped
regime does not hold well in pure solvent but becomes satisfactory with the increase of
viscosity. A generalization of the
theory with taking into account the inertial term in the
ordinary Langevin equation is indispensable for the description of reactions in
a gas phase that proceed in the underdamped limit.
Those in condensed media (in solution or in an enzyme) are generally believed to
proceed typically in the overdamped regime and we also resort to this assumption
for simplification of further analysis. We stress once more the reason why
for the motion of the peptide group we apply the underdamped regime while
for the movement along the reaction coordinate we invoke to the overdamped
one. In the former case the linear displacements of the atoms are
$\approx 0.1\ \AA$, i.e., much less than the size of the solvent molecules
 $\approx 1\ \AA$ (see Sec. 2). In the latter case the
linear displacements of atoms are
comparable with the size of the solvent molecules. That is why the friction
is supposed to be sufficiently strong for the overdamped regime to be
applicable.

Most important of all we resort to a severe approximation based on the
following reasoning. Strictly
speaking a Fourier mode of the random process $\phi(t)$
\begin{equation}
\label{eq26} \tilde \phi(\omega)=\frac{1}{2\pi}\int_{-\infty}^{\infty}dt\
\phi(t) exp(-i\omega t)
\end{equation}
is a random function of frequency with zero mean. However the mean squared
amplitude of this mode is related to the correlation function $\alpha(t)$
via the Wiener-Khinchin theorem
\begin{equation}
\label{eq27} <\mid \tilde \phi(\omega)\mid^2>=\int_{-\infty}^{\infty}dt\
\alpha(t) exp(-i\omega t)=2\pi \tilde \alpha(\omega)
\end{equation}
where $\tilde \alpha(\omega)$ is given by (\ref{eq10}).
We can introduce the effective mean amplitude at any frequency as the square
root of (\ref{eq27}). In particular we can do it for the frequency $\omega_0$
\begin{equation}
\label{eq28} <\mid \tilde \phi(\omega_0)\mid>_{eff}=
\sqrt{2\pi \tilde \alpha(\omega_0)}
\end{equation}
As was already mentioned after (\ref{eq10}) the spectrum of $\alpha(\omega)$
at the requirement (\ref{eq7}) has a sharp resonant peak at the frequency
$\omega_0$. It means that practically the whole power of the spectrum is in
the Fourier mode $\tilde \phi(\omega_0)$ of the random process $\phi(t)$
at this frequency. That is why we can roughly approximate the random process
$\phi(t)$ by a harmonic vibration with the amplitude given by (\ref{eq28})
\begin{equation}
\label{eq29} \phi(t)\approx <\mid \tilde \phi(\omega_0)\mid>_{eff}\
sin(\omega_0 t)
\end{equation}
From (\ref{eq9}), (\ref{eq10}) with taking into account (\ref{eq7}) we obtain
\begin{equation}
\label{eq30} \phi(t)\approx \sqrt{\frac{\sqrt 2 k_BT}{\gamma \omega_0}}
\ sin(\omega_0 t)
\end{equation}
This artificial replacement of the oscillating noise
(that is a random process with the properties (\ref{eq25}))
 by oscillations (\ref{eq30})  simplifies the problem significantly and
enables us to employ some known results from the Kramers' theory. As was
stressed above it is a
methodical trick enabling us to cast the problem into a tractable form
rather than an indespensable assumption for our approach.

\section{Effect of the RPV}
A standard tool to investigate dynamical effects in  reaction rate is
the Kramers' theory (for review see  \cite{Han90}, \cite{Jun93},
\cite{Fle93}, \cite{Tal95}
  and refs. therein).
The latter is based on the ordinary Langevin equation of motion along the
reaction coordinate $q$.
In the Kramers' theory a chemical reaction is modeled as the escape
of a Brownian particle with the mass $m$
(the effective mass of the reaction coordinate, e.g., the reduced mass of
 a scissile bond) and linear size $l$ in the
media with viscosity $\eta$ (so that the friction coefficient is given by
the Stokes formula $\nu=6\pi l \eta$) and temperature $T$
from the well of a
metastable potential along the reaction coordinate.
This potential is an electronically adiabatic
ground state obtained in the Born-Oppenhaimer approximation
 by methods of quantum chemistry. It is considered as preliminary input
information for the Kramers' theory and neither its origin nor its possible
modification by catalysts is an issue of the present approach.

For the potential $V(q)$ we consider an arbitrary form with a
barrier that has at the top of the latter the value
$V^{\prime\prime}(q_b)\approx -m\omega^2_b$ and at the bottom
of the well the value $V^{\prime\prime}(q_a)\approx m\omega^2_a$.
Also we include into consideration the oscillating force
\begin{equation}
\label{eq31} g\phi(t)=h\ sin(\omega_0 t)
\end{equation}
where
\begin{equation}
\label{eq32} h=\frac{2ep}{\epsilon R^3}
\sqrt{\frac{\sqrt 2 k_BT}{\gamma \omega_0}}
\end{equation}

As a result the ordinary Langevin equation of motion along the reaction
coordinate $q$ (\ref{eq24}) takes the form
\begin{equation}
\label{eq33} \nu\frac{dq}{dt}=
-V^{\prime}(q)+h\ sin(\omega_0 t)+\zeta (t)
\end{equation}
We introduce the dimensionless variables and parameters as follows
\begin{equation}
\label{eq34}  q\rightarrow q (q_b-q_a);\ \ \ \ \
t\rightarrow t\frac{\nu}{m\omega^2_b};
\ \ \ \ \  \omega_0\rightarrow \frac{\omega_0 \nu}{m\omega^2_b}=\Omega
\end{equation}
and denote
\begin{equation}
\label{eq35} A=\frac {h}{m\omega^2_b(q_b-q_a)};
\ \ \ \ \ D=\frac{k_B T}{m\omega^2_b(q_b-q_a)^2}
\end{equation}
where $A$ is the dimensionless parameter characterizing the
oscillating electric field strength and $D$ is the dimensionless
parameter characterizing the intensity of thermal fluctuations
(i.e., the temperature of the heat bath) relative to the barrier
height. Their ratio is the most important parameter of the model
\begin{equation}
\label{eq36} \frac{A}{D}=\frac{h(q_b-q_a)}{k_B T}=\frac{2ep(q_b-q_a)}{\epsilon R^3}
\sqrt{\frac{\sqrt 2 }{\gamma \omega_0 k_B T}}
\end{equation}
In dimentionless variables the Langevin equation is
\begin{equation}
\label{eq37} \frac{dq}{dt}=-U'(q)+A sin(\Omega t)+
\chi(t)
\end{equation}
where $U(q)$ is the dimensionless potential and the dimensionless noise
 $\chi(t) =
\frac{1}{m\omega_b^2(q_b-q_a)}
\zeta\left(\frac{t\nu }{m\omega_b^2}\right)$ is characterized by
$<\chi(t)>=0$ and $<\chi(0)\chi(t)>=2D\delta(t)$.
The corresponding overdamped limit of the  Fokker-Planck
equation for the probability
distribution function $P(q,t)$ \cite{Han90} is
\begin{equation}
\label{eq38} \frac{\partial P(q,t)}{\partial t}=-\frac{\partial
}{\partial q}\Bigl\lbrace [-U'(q)+A sin(\Omega
t)]P(q,t)\Bigr\rbrace+D\frac{\partial^2P(q,t)}{\partial q^2}
\end{equation}

In the absence of driving ($A=0$) the escape rate is given by
the famous Kramers' formula
\begin{equation}
\label{eq39} \Gamma_K
\approx \frac{\omega_a \omega_b}{2\pi }
exp\Bigl[-\Bigl(U(q_b)-U(q_a)\Bigr)/D\Bigr]
\end{equation}
where $\omega_a=\sqrt{U^{\prime\prime}(q_a)}$ and
 $\omega_b=\sqrt{\vert U^{\prime\prime}(q_b)\vert}$.
In the case of thermally activated escape with periodic driving
($A \not= 0$) one usually introduces
 the instantaneous rate constant $\Gamma(t)$, the rate constant
$\overline\Gamma$ averaged over the
 period of oscillations $T=2\pi/\Omega$
and is interested in the escape rate enhancement
 $\overline\Gamma/\Gamma_K$ (see \cite{Jun93},
\cite{Leh000}, \cite{Sme99}, \cite{Dyk051} and refs. therein).
Regretfully no workable formula for the general case of arbitrary
modulation amplitude to noise intensity ratio $A/D$ and arbitrary frequency
$\Omega$ is available at present. However the case of moderately weak to
moderately strong modulation $ D << A << \sqrt D$ is relevant for our problem.
In this case a simple formula
for the escape rate enhancement is known \cite{Sme99}
\begin{equation}
\label{eq40} \Delta\equiv
\frac{\overline\Gamma}{\Gamma_K}\approx\frac{\sqrt D}{
\sqrt{2\pi A c\left(\Omega \right)\left(1+O(A)\right)}}
exp\Biggl [ \frac{A c\left(\Omega \right)}{D}
\left(1+O(A)\right)\Biggr]
\end{equation}
where for the
cubic (metastable) potential
$c\left(\Omega \right)=\pi \Omega/sh\left(\pi \Omega\right)$  and
for the quartic (bistable) potential
$c\left(\Omega \right)=\sqrt {\pi \Omega/sh\left(\pi \Omega\right)}$.
This formula is sufficient for
the purposes of the present paper to understand the solvent viscosity
dependence in the preexponent. However it should be stressed that it is
invalid is one wish to evaluate reaction rate enhancement due to the
suggested mechanism. In this case the corrections $O(A)$ lead to
deviation from the log-linear behavior predicted by (\ref{eq40}).
Fortunatly the latter problem is not a matter of the present paper and
(\ref{eq40}) suffices for our needs. However we briefly return to the
problem of evaluation of the reaction rate enhancement due to the suggested
mechanism in the next Sec. and in Appendix B.

\section{Numerical estimates}
Before proceed further we should define the range of the
parameters for our model. To be supported by evidence we take the
numerals for a very typical and one of the most studied enzymatic
reaction catalysed by Subtilisin. This enzyme belongs to serine
proteases and brakes the bond between the atoms C and N in a
peptide group of a substrate of protein nature. At physiological
temperatures where enzymes normally work ($k_B T \approx
4.2\cdot10^{-14}\ CGS$) the corresponding noncatalysed reaction
in solution typically has the rate constant $k\approx 1 \cdot
10^{-8}\  s^{-1}$ while the enzymatic reaction has the
rate constant (that of the rate limiting step) $k_{cat}\approx 5
\cdot 10^1\  s^{-1}$ \cite{Car88}. Thus the total catalytic
effect is approximately $10^{12}$. We have for the friction
coefficient of the movement along the reaction coordinate
$\nu=6\pi l \eta$ at $l\approx 1\ \AA$ and $\eta\approx 1\ cP$
(that is the viscosity of water at room temperature) the estimate
 $\nu\approx 2\cdot 10^{-9}\ CGS$.
Then taking into account that we measure dimensionless time in the units of
$\frac{\nu}{m\omega^2_b}$ (see (\ref{eq34}))
we have for the reaction rate constant of the noncatalysed reaction in dimensionless
form at $m\approx 5\ a.u.m.$ ($a.u.m.\approx 1.7 \cdot 10^{-24} g$)
and $\omega_b\approx 10^{14} s^{-1}$ the
value $\Gamma=\frac{k \nu}{m \omega_b^2}\approx 10^{-22}$. We can
evaluate from the Arrhenius factor of the Kramers' rate in the dimensionless
form  for the quartic (bistable) potential $V(q)=-q^2/2+q^4/4$  \cite{Han90}
\begin{equation}
\label{eq41} \Gamma_K =(\sqrt 2 \pi)^{-1} exp\biggl[-1/(4D)\biggr]
\end{equation}
that in order to obtain such a reaction rate constant in the
absence of the oscillating electric field ($A=0$) we should take
$D\approx \frac{1}{4\cdot 22\  ln 10} \approx 5\cdot 10^{-3}$. If we
take the cubic (metastable) potential $V(q)=q^2/2-q^3/3$ then we obtain
 $D\approx  3\cdot 10^{-3}$. Thus the typical noise intensity for
enzymological problems is
\begin{equation}
\label{eq42}  D\approx  3\div 5 \cdot 10^{-3}
\end{equation}
As earlier we assume
that the dimensional frequency of the RPV is on the picosecond time scale and is
approximately $\omega_0\approx 100\ cm^{-1}\approx 10^{13}\ s^{-1}$. Then from
(\ref{eq34}) we obtain
at $m\approx 5\ a.u.m.$ and $\omega_b\approx 10^{14}\ s^{-1}$
that the dimensionless frequency in our case is
\begin{equation}
\label{eq43} \Omega \approx 2 \cdot 10^{-1} << 1
\end{equation}
Thus we are actually in the low frequency regime
or that of slow modulation $\Omega << 1$.
In this case we have $c\left(\Omega \right)\approx 1$ for both cubic
(metastable) and quartic (bistable) potentials. That is why
$c\left(\Omega \right)$ will be
droped further from the formula for the reaction rate enhancement.

We take the value $p=3.6\ D$ and typical values: for the partial charge
$e\approx 0.3\ a.u.c.$ ($a.u.c.=4.8 \cdot 10^{-10}\ CGS$), for the reaction potential
$q_b-q_a\approx 1\ \AA$, for the dielectric constant in protein interior
$\epsilon\approx 3.5$, for the distance of the fluctuating peptide group from
the reaction coordinate $R\approx 3\ \AA$, for the frequency of the
RPV $\omega_0\approx 100\ cm^{-1}\approx 10^{13}\ s^{-1}$ and for the
friction coefficient of the peptide group (see Sec.2)
 $\gamma \approx 2\cdot 10^{-27}\ CGS$. Then at
room temperature we obtain from (\ref{eq36})
that $A/D \approx 10$. At such value of this parameter the formula (\ref{eq40})
predicts the reaction rate enhancement by $3 \div 4$ orders of magnitude
($log\ \Delta\approx 3.4$). This estimate requires comments. As was stressed above the
formula (\ref{eq40}) overestimates the reaction rate enhancement. As a
matter of fact the corrections $O(A)$ lead to some more moderate growth than
the log-linear one \cite{Leh000}, \cite{Sme99}, \cite{Dyk051}. However
the results presented in these papers testify that an appreciable
deviation from the log-linear behavior begins at significantly higher values of
modulation amplitude to noise intensity ratio $A/D$ than $A/D \approx 10$. That is
why the reaction rate enhancement by $\approx 3 \div 4$ orders of
magnitude at $A/D \approx 10$ due to the suggested  mechanism
seems to be quite feasible. Here we restrain ourselves from further
discussing this mostly important issue because the present paper is devoted
to a particular effect manifesting itself in the preexponent rather than in
the exponent of the formula (\ref{eq40}). However in Appendix B we return to
this problem and discuss it in more details.

\section{Solvent viscosity effect}
Returning to dimensional parameters, substituting (\ref{eq36}) and
(\ref{eq39}) into (\ref{eq40}) and dropping
$c\left(\Omega \right ) \approx 1$ (see previous Sec.) we obtain for
the reaction rate constant
\[
 \bar \Gamma=\Gamma_K \Delta\approx
\frac{m \sqrt{V^{\prime\prime}(q_a)
\vert V^{\prime\prime}(q_b)\vert}}{2 \pi \nu}
exp\left[-\frac{\Bigl(V(q_b)-V(q_a)\Bigr)}{k_B T}
\right]\times
\]
\begin{equation}
\label{eq44}\sqrt{\frac{\epsilon R^3}{4\pi ep(q_b-q_a)}
\sqrt{\frac{\gamma \omega_0 k_B T}{\sqrt 2}}}
exp\Biggl [ \frac{2ep(q_b-q_a)}{\epsilon R^3}
\sqrt{\frac{\sqrt 2 }{\gamma \omega_0 k_B T}}\Biggr ]
\end{equation}
First we consider the hypothetical "ideal" cosolvent with infinitely
small molecular weight
(cosolvent molecules freely penetrate into enzyme and are distributed
there homogeneously).
Taking into account that in this case we have both
$\nu \propto \eta$ ($\nu=6\pi l \eta$) and $\gamma \propto \eta$
(see (\ref{eq6})) we obtain
\begin{equation}
\label{eq45}  \bar \Gamma \propto \frac{1}{\eta^{0.75}}
\end{equation}
Thus the present approach yields the limit value $\beta_{max}= 0.75$
for the "ideal" cosolvent that is rather close to the extrapolated limit value
$\beta_{max}\approx 0.79$  from the paper \cite{Yed95}.
The dependence on $\eta$ in the second exponent is strongly overshadowed
by the first exponent because typically we have
\begin{equation}
B=\label{eq46} \frac{V(q_b)-V(q_a)}{k_B T} >>
b=\frac{2ep(q_b-q_a)}{\epsilon R^3}
\sqrt{\frac{\sqrt 2 }{6\pi l \omega_0 k_B T}}
\end{equation}
Then in the $ln$ of the reaction rate constant
$ln\ \bar \Gamma=const-0.75\ ln\ \eta + B -b/\sqrt {\eta}$ the last term is
negligible provided the requirement (\ref{eq46}) is satisfied.

In the $log\ \bar \Gamma$ vs. $log\ \eta$ coordinates the formula
(\ref{eq44}) yields the lines that resemble the behavior of
the straight ones from
the paper \cite{Yed95}. To exhibit it explicitly let us return to
dimensionless variables and parameters and write $\eta =\eta_0\ 10^x$ where
$\eta_0=1\ cP$ at which we typically have $\left(A/D\right)_0 \approx 10$
(see previous Sec.). Then we obtain
\begin{equation}
\label{eq47} log\ \bar \Gamma = log \left[ \frac{\omega_a \omega_b}{2\pi}\sqrt{\frac{1}
{2\pi}\left(\frac{D}{A}\right)_0}\right]-
\frac{log\ e}{nD}-\frac{3x}{4}+\left(\frac{A}{D}\right)_0\ \frac{log\ e}{10^{x/2}}
\end{equation}
where for the cubic (metastable) potential $\omega_a = 1$, $\omega_b = 1$
and $n=6$ while
for the quartic (bistable) potential $\omega_a = \sqrt 2$, $\omega_b = 1$
and $n=4$. The plots for the latter case obtained with the help of
 the formula (\ref{eq47})
are depicted in Fig.5. However we should recognize that the lines obtained
are in bad accordance with the results of the paper \cite{Yed95}. At
realistic values $A/D \approx 10$ we obtain too large decrease of the
reaction rate constant $\approx 6$ orders of magnitude instead of $\approx 1$ order
of magnitude in \cite{Yed95} at the increase of solvent viscosity by $2$ orders of magnitude.
Besides the lines for this case deviate appreciably from the straight ones obtained
in \cite{Yed95}.
That is why in Appendix B we return to this problem and show that taking into account
the corrections $O(A)$ to the formula (\ref{eq40}) for the reaction rate
enhancement improves the situation significantly and yields the results that are in
good agreement with experimental data from the paper \cite{Yed95}.

Now we consider a realistic cosolvent with finite molecular weight $M$. In this
case both friction coefficients $\nu$ and $\gamma$ are some unknown
functions of solvent
viscosity.  The cosolvent molecules with larger molecular weight worser
penetrate both into the enzyme active site where the reaction coordinate is
located and into the enzyme body where the functionally important
cranckshaft-like motion of favourably located peptide groups takes place.
That is why with the increase of $M$ these functions should grow more slowly
the higher the $M$ is. Thus cosolvent molecules with larger molecular weight
should create less pronounced dependence on solvent viscosity than those
with lesser molecular weight. In our opinion there are not enough
experimental data to define these functions reliably on physical grounds.
That is why we further resort to speculations of purely phenomenological
character. We assume that these functions are of the form
$\nu \propto \eta^{\delta (M)}$ and $\gamma \propto \eta^{\alpha (M)}$
with the requiremens that the indecies of a power $\delta (M)$ and
$\alpha (M)$ obey $\lim_{M\to 0}{\delta (M)} = 1$,
$\lim_{M\to 0}{\alpha (M)} = 1$ and are decaying
functions with the increase of $M$. These requirements take into account
the above mentioned features.

Then the formula (\ref{eq44}) yields
\begin{equation}
\label{eq48}  \bar \Gamma \propto \frac{1}{\eta^{\beta (M)}}
\end{equation}
where
\begin{equation}
\label{eq49}  \beta (M)= \delta (M)-\frac{\alpha (M)}{4}
\end{equation}
Regretfully the experimental data available for $\beta (M)$  do not
allow us to distinguish $\delta (M)$ and $\alpha (M)$ separately.
We can only say that from the requirement $\beta (M) > 0$ we should have
$\delta (M) > \alpha (M)/4$. The
dependence of the parameter $\beta$ on the values of $\delta$ and $\alpha$
is presented in Fig. 6. One can distinguish two regions (see also Fig. 7):
in the region $I$ we have $\beta > 0.75$ while in the region $II$ we have
$\beta < 0.75$. Accordingly three possible types of behavior are
concievable.
We call them normal, rare and exceptional and schematically depict in Fig.7.
Let us start from the case of large cosolvent molecular weight when both
$\delta (M)$ and $\alpha (M)$ are small. For the rare type of behavior
the decrease of $M$ leads to a path in the $\alpha - \delta$ parameter space
that is always left in the region $I$, i.e., the value $\beta (M)$ is always
less than the limit value $\beta_{max}=0.75$. For the normal type of
behavior the path crosses the fat line and enters the region $II$ where
 $\beta (M) > 0.75$. At the minimal realistic cosolvent molecular weight
$M=62$ for Ethylenglycol (see \cite{Yed95}) we have some $\alpha_n^{min}$
and $\delta_n^{min}$ at which we generally can obtain arbitrary
$0.75 < \beta_{max} < 1$. Finally for the exceptional type of behavior we can
obtain $\alpha_e^{min}$ so small and $\delta_e^{min}$ so large that
$\beta_{max}=\delta_e^{min}- \alpha_e^{min}/4 \approx 1$, i.e., the
approximate dependence $\bar \Gamma \propto 1/\eta$ emerges.

Also the formula (\ref{eq44}) yields the dependence of the reaction rate
constant on dielectric constant $\epsilon$. Such dependence has been known from
the experiment for a long time \cite{Gav79} and is of the type
$ln\ k_{cat} = ...+const/\epsilon$. The formula (\ref{eq44}) predicts
$ln\ \bar \Gamma=...+const/\epsilon+(1/2)ln\ \epsilon$. It is obvious that
the latter term is practically unnoticeable on the background of the former
one. That is why one can conclude that the formula (\ref{eq44}) yields
correct dependence on dielectric constant in accordance with experimental
data from \cite{Gav79}.

\section{Conclusions}
We suggest a dynamical mechanism that mediates the influence of solvent
viscosity on the reaction coordinate at enzyme action. The mechanism is
based on the fluctuations of the electric field in the enzyme active
site. These fluctuations are produced by thermally equilibrium dynamics of
protein structure, namely by rocking of the rigid plane of a favourably located
and oriented peptide group (or may be a few of them). Such rocking causes
the electric field of the dipole moment of the peptide group lying in its
plane to undergo fluctuations in time. As the latter are thermally
equilibrium they exist on the whole time scale of enzyme turnover and there
is no problem to match them with the catalytic act. Namely the impossibility
of survival for artificially constructed coherent oscillatory excitations in
protein dynamics on the
time scale of the catalytic act plagues many of the
suggestions for dynamical contribution into enzyme action and causes
justifiedly criticism of Warshel and coauthors \cite{Ols06}, \cite{Ols006},
 \cite{War06}. Such excitations may really be created
by energy released at substrate binding by an enzyme.
They can exist for some life time
in the form of, e.g., the so called discrete breathers either in
protein highly regular secondary structure \cite{Sit06}, \cite{Sit07}
or in its whole irregular tertiary structure \cite{Jua07}.
However a protein is
condensed media and any motion both in its interior and on its surface
of thermally nonequilibrium characher must fade away rapidly
(on the enzyme turnover time scale) due to dissipation.
Long life times for discrete breathers obtained in \cite{Jua07}
 are due to unphysical
assumption of friction for only surface elements of the
three-dimensional network of
oscillators. In our opinion the only possibility is to obtain some
kind of influence on the reaction coordinate (having no counterpart for
 reactions in solution)
from thermally equilibrium fluctuations. The structure of the protein
enables peptide groups to implement high
frequency thermally equilibrium rocking of their rigid planes
(cranckshaft motion).
This is a rotational motion and linear displacements of the atoms at it are
negligibly small compared with both the size of the solvent molecules and
the interatomic distances in protein. That is why this type of
motion proceeds in the underdamped regime even in the condensed media of
protein interior and mimics oscillations very much.

The specific Fourier mode of the electric field fluctuations at some
 own frequency of rocking of the rigid plane of the peptide group
(presumably $\omega_0 \sim 100\ cm^{-1}$) resembles coherent
 oscillation and can be
approximated by a harmonic vibration. The latter is the RPV in
our approach. The rocking of the rigid plane of the peptide group feels the
microviscosity of the environment in the vicinity of the latter.
This microviscosity is a
function of solvent viscosity. Details of this function depend on the
molecular weight of the cosolvent very much. For the hypothetical "ideal"
cosolvent with infinitely small molecular weight
(cosolvent molecules freely penetrate into enzyme and are distributed
there homogeneously) the microviscosity is identical to solvent viscosity.
For realistic cosolvent their relationship is more complicated and
poorly known.

The mechanism suggested in the
present paper yields the required functional dependence (\ref{eq1})
of an enzymatic reaction rate constant on solvent viscosity and the
limit value $\beta_{max}\approx 0.75$ that is in good agreement with the
extrapolated one  $\beta_{max}\approx 0.79$  from the paper \cite{Yed95}
(see Introduction).
We have considered predominantly the limiting case of the hypothetical "ideal"
cosolvent. To take into account the realistic molecular weight of
the cosolvent one needs to know how its molecules are located in the enzyme
 both relative
to the reaction coordinate and the rocking peptide group (or a few of them)
interacting with
the latter via electric field fluctuations.
In this case both the friction coefficient for the reaction coordinate
$\nu$ and that for the rocking peptide group $\gamma$ become more complicated
functions of solvent viscosity $\eta$ than simple proportionality.
All the arguments from the
papers \cite{Gav79}, \cite{Bee80}, \cite{Dos83}, \cite{Gav86},
\cite{Gav94}, \cite{Yed95}, \cite{Bar95}, \cite{Kle98}
remain pertinent for this case. However we do not
feel that there are enough data at present to tackle realistic cosolvent
within our approach at a deeper level than purely phenomenological speculations.
We are merely convinced that it should be done not
within the Kramers' formula yielding $k_{cat}\propto 1/\eta$ dependence
but within a formula of the type of (\ref{eq44}) where the upper limit
$\beta_{max}$
is already built-in, i.e., the dependence $k_{cat}\propto 1/\eta^{\beta}$ with
$\beta \leq \beta_{max}$  is obtained automatically.

Regretfully at present we have to work with very limited tools from the Kramers'
theory. The most natural formulation of our problem leads to a two noises
model one of which is the oscillating noise (\ref{eq25}).
There are no results for such model in the literature yet.
Further development of the model will inevitably require its elaboration
within the approach of the papers \cite{Ray01},
 \cite{Ray07}. At present stage we have to resort to artificial
approximating  the oscillating noise by coherent oscillations.
However even for the latter simplified case the
theory of the Fokker-Planck equation (\ref{eq38})
 does not provide us at present with a
workable formula for the general set of the parameters $A$, $D$ and
$\Omega$. This fact does not enable us to
evaluate reliably "how much can the electric field fluctuations in the
enzyme active site contribute into the reaction rate enhancement ?"
 because small corrections in the exponent can lead to drastic
overestimations. However we argue that the low frequency regime
is relevant for enzymological problems. For this case the corrections $O(A)$
to the formula (\ref{eq40}) can be obtained explicitly (see Appendix B).
They allow us conclude that the reaction rate enhancement by the suggested
mechanism up to $\approx 3 \div 4$ orders of magnitude is feasible.
We argue that available results enable us to consider the preexponent quite
safely. As the functional dependence (\ref{eq1})
of an enzymatic reaction rate constant on solvent viscosity manifests itself
namely in the preexponent the employed formula (\ref{eq40}) seems to be
sufficient for the purposes of the present paper. The discrepansies with
experimental data are eliminated by taking into account the corrections $O(A)$
to this formula as the results from Appendix B testify.

We conclude that the present
approach grasps the main feature of the functional dependence
(\ref{eq1}) for enzymatic reaction rate constant on solvent viscosity
 and satisfactorily yields the upper limit for the fractional
index of a power in it.

\section{Appendix A}
Here we give a brief estimate of the moment of inertia of the peptide group
for rotation around the axis $\sigma$ that goes through the
center of masses of the peptide group parallel to the bonds
$C_{\alpha}^{i-1}-C^i$ and $N^i-C_{\alpha}^{i}$ (see Fig.1).
The peptide group is a rigid plane structure so that all atoms $O$, $C$, $N$ and $H$
lie in a plane. That is why the moment of inertia is simply the sum of their
masses multiplied by the square of their distances to the axis $\sigma$.
The masses in atomic units ($1\ a.u.m.=1.7 \cdot 10^{-24} g$) are $m_O=16$,
$m_C=12$, $m_N=14$ and $m_H=1$. That is why the center of masses is shifted
a little to the atoms $C$ and $O$ relative the middle of the bond $C-N$. The
lengths of the bonds are $l_{OC}=1.24\ \AA$, $l_{CN}=1.32\ \AA$ and
 $l_{NH}=1.\ \AA$. That is why the distances to the axis $\sigma$ are
approximately  $r_{O}\approx 1.5\ \AA$, $r_{C}\approx 0.3\ \AA$ and
 $r_{N}\approx 0.7\ \AA$ and $r_{H}\approx 2\ \AA$. Substituting these values into the
formula $I=\sum m_i r_i^2$ we obtain  $I \approx 7.6 \cdot 10^{-39}
g \cdot cm^{2}$. Thus our rough estimate corroborate the  precise value
 $I \approx 7.34 \cdot 10^{-39}
g \cdot cm^{2}$ from \cite{Flu94}.

\section{Appendix B}
We consider the low frequency regime $\Omega << 1$ that is argued
in Sec. 6 to be relevant for our problem. For the particular case of
moderately strong modulation $A << A_c$ (where the bifurcational point
$A_c$ has the values $A_c =0.25$ for
the case of the cubic (metastable) potential $U(q)=q^2/2-q^3/3$ and
$A_c = 2/(3\sqrt 3)\approx 0.4$ for
the case of the quartic (bistable) potential $U(q)=-q^2/2+q^4/4$) the reaction
rate enhancement is obtained in \cite{Sit061}. The formula obtained there is
rather cumbersom and is not presented here to save room. The results
obtained with its help for the case of quartic (bistable) potential are
depicted in Fig.8. They testify that at our values $D = 5 \cdot 10^{-3}$ and
$A/D \approx 10$ (see Sec. 6) the reaction rate enhancement is $\sim 3 \div 4$
 orders of magnitude. Thus the conclusion that the suggested mechanism is
strong enough to affect enzyme action appreciably is justified.
For the intermediate regime of moderately weak to
moderately strong modulation
$D << A << \sqrt D$ (our estimate
 $A/D \approx 10$ is within this range for practically important for enzymology
range $ D\approx  3\div 5 \cdot 10^{-3}$ (\ref{eq42})) the formula is simplified
significantly and yields an explicit expression for the corrections $O(A)$
to the formula (\ref{eq40}). We denote
\begin{equation}
\label{eq50} p =
1+\frac{2}{\omega_b^2}\Biggl[
\frac{\omega_a^2}{4}+\frac{\pi-2}{\pi}
\Biggl( \frac{1}{2\omega_b}-\frac{1}{\pi}\Biggr)
\Biggr ]
\end{equation}
where for the cubic (metastable) potential (CP) $\omega_a=\sqrt{U^{\prime\prime}(q_a)}= 1$,
$\omega_b=\sqrt{U^{\prime\prime}(q_b)} = 1$ and $q_a=0$, $q_b=1$ while
for the quartic (bistable) potential (QP) $\omega_a =\sqrt{U^{\prime\prime}(q_a)}= \sqrt 2$,
 $\omega_b=\sqrt{U^{\prime\prime}(q_b)} = 1$ and $q_a=-1$, $q_b=0$.

Then we have a simple formula for the reaction rate enhancenment
\begin{equation}
\label{eq51} \Delta \approx
\frac{\sqrt D}{2
\sqrt{\pi A\Bigl[\left(q_b-q_a\right)/2-
A p\Bigr]}}
exp\Biggl \{ \frac{A}{D}\Bigl[q_b-q_a -
A p\Bigr]\Biggr \}
\end{equation}
where $p$ is the constant given by (\ref{eq50}).
For the case of CP we have $p\approx 1.632$
while for the QP we have  $p\approx 2.132$. For both of them we have
$q_b-q_a = 1$.

From (\ref{eq51}) we obtain a modification of the formula (\ref{eq47})
\[
 log\ \bar \Gamma = log \left[ \frac{\omega_a \omega_b}{2\pi}\right]+
log \left[ \frac{1}{2\sqrt \pi} \sqrt{\left(\frac{D}{A}\right)_0}\right]-
\frac{log\ e}{nD}-\frac{3x}{4}-
\]
\begin{equation}
\label{eq52} \frac{1}{2}log \left[ \frac{1}{2}-
p D \left(\frac{A}{D}\right)_0\ \frac{log\ e}{10^{x/2}}\right]+
\left(\frac{A}{D}\right)_0\ \frac{log\ e}{10^{x/2}}
\left[ 1-p D \left(\frac{A}{D}\right)_0\ \frac{log\ e}{10^{x/2}}\right]
\end{equation}
The results obtained with the help of this formula for the case of quartic
(bistable) potential ($\omega_a = \sqrt 2$, $\omega_b = 1$ and $n=4$) are
depicted in Fig. 9. They are good agreement with the results of the paper
\cite{Yed95} because they yield straight lines for the decrease of the
rate constants by approximately one order of magnitude at the increase
of solvent viscosity by two orders of magnitude.\\

Acknowledgements. The author is grateful to Dr. Yu.F. Zuev
for helpful discussions. The work was supported by
the grant from RFBR.

\newpage

\begin{figure}
\begin{center}
\includegraphics* [width=\textwidth] {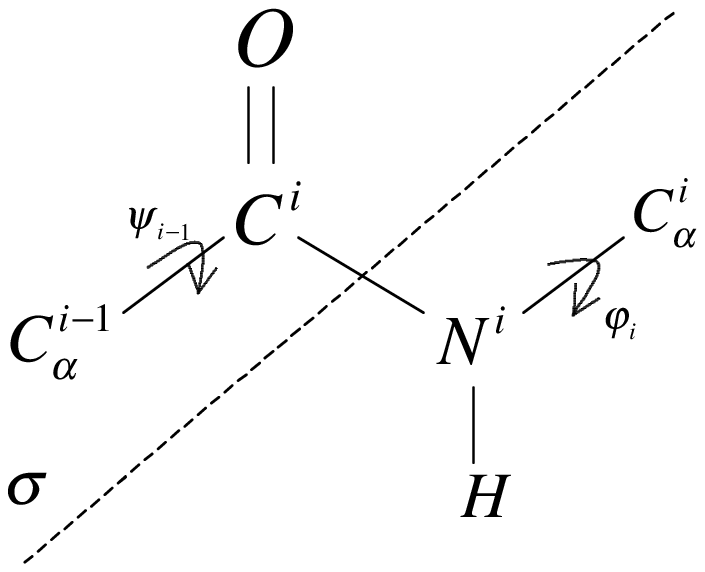}
\end{center}
\caption{Schematic picture of the peptide group.} \label{Fig.1}
\end{figure}

\clearpage
\begin{figure}
\begin{center}
\includegraphics* [width=\textwidth] {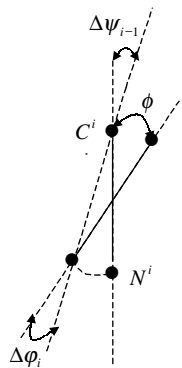}
\end{center}
\caption{A look on the peptide group from the axis of rotation $\sigma$
explaining the definition of the angle $\phi$ (defined in
(\ref{eq3})).}
\label{Fig.2}
\end{figure}

\clearpage
\begin{figure}
\begin{center}
\includegraphics* [width=\textwidth] {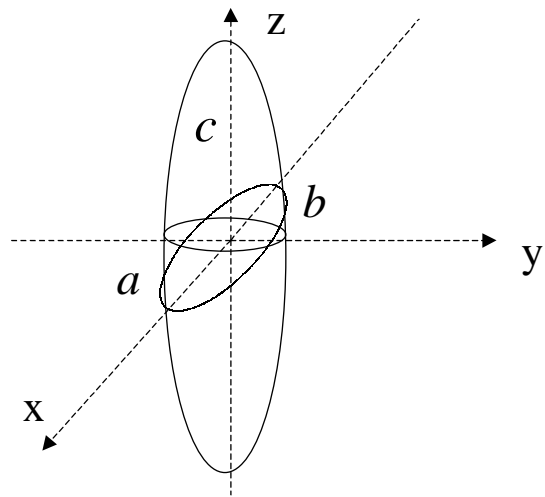}
\end{center}
\caption{Model of the peptide group by oblate ellipsoid.}
\label{Fig.3}
\end{figure}

\clearpage
\begin{figure}
\begin{center}
\includegraphics* [width=\textwidth] {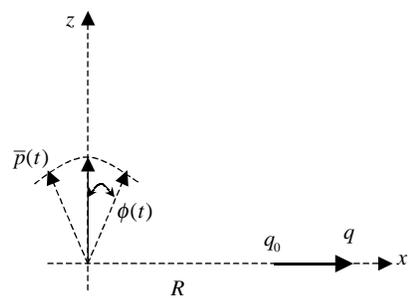}
\end{center}
\caption{Favourable orientation of the peptide group
 dipole moment $\bar p(t)$ relative to the reaction coordinate $q$.}
\label{Fig.4}
\end{figure}

\clearpage
\begin{figure}
\begin{center}
\includegraphics* [width=\textwidth] {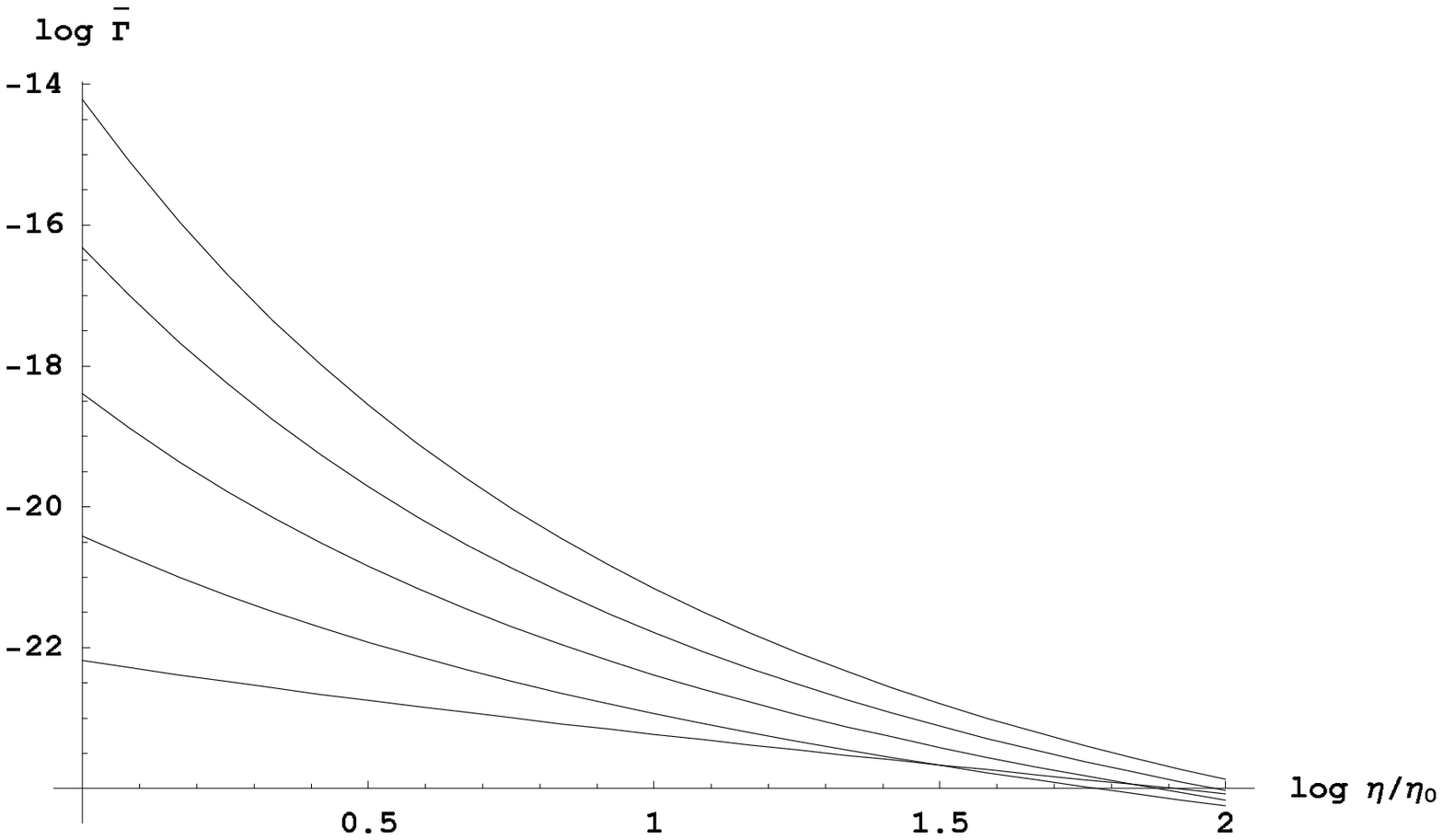}
\end{center}
\caption{Dependence of the reaction rate constant $\bar \Gamma$
 on solvent viscosity $\eta$  in the $log\ \bar \Gamma$ vs.
$log\ \left(\eta/\eta_0\right)$
coordinates given by formula (\ref{eq47}) for the case of quartic (bistable) potential.
The values of $\left(A/D\right)_0$ from the down
line to the upper one respectively are: 1, 6, 11, 16, 21.}
\label{Fig.5}
\end{figure}

\clearpage
\begin{figure}
\begin{center}
\includegraphics* [width=\textwidth] {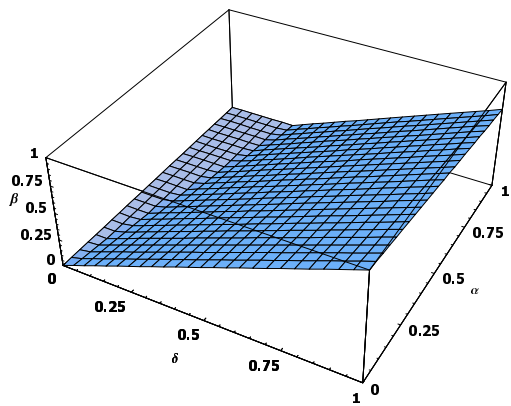}
\end{center}
\caption{Dependence of the parameter $\beta=\delta-\alpha/4$ (see (\ref{eq48}))
on the values of $\delta$ and $\alpha$.}
\label{Fig.6}
\end{figure}

\clearpage
\begin{figure}
\begin{center}
\includegraphics* [width=\textwidth] {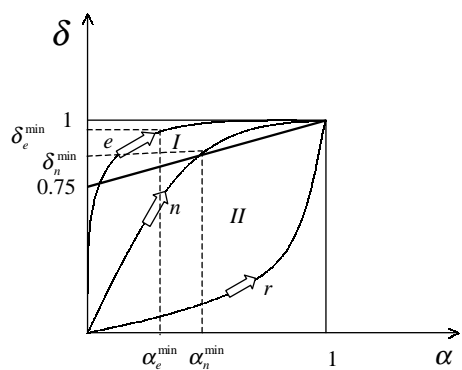}
\end{center}
\caption{Three types of possible behavior of the parameters
$\delta (M)$ and $\alpha (M)$ on cosolvent molecular weight $M$: $n$-normal,
$r$-rare and $e$-exceptional. The fat line devides the parameter space to the region
$I$ where $\beta=\delta-\alpha/4 > 0.75$ and the region $II$ where
$\beta=\delta-\alpha/4 < 0.75$.}
\label{Fig.7}
\end{figure}

\clearpage
\begin{figure}
\begin{center}
\includegraphics* [width=\textwidth] {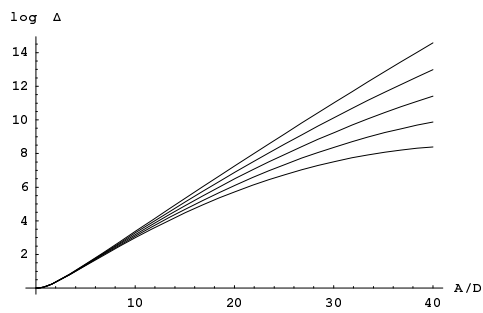}
\end{center}
\caption{The dependence of the escape rate enhancement on
the driving amplitude to noise intensity ratio for the case of
quartic (bistable) potential $U(q)=-q^2/2+q^4/4$. The values
of the noise intensity $D$ from the down
line to the upper one respectively are:
  $5\cdot 10^{-3}$; $4\cdot 10^{-3}$; $3\cdot 10^{-3}$;
$2\cdot 10^{-3}$; $1\cdot 10^{-3}$.}
\label{Fig.8}
\end{figure}

\clearpage
\begin{figure}
\begin{center}
\includegraphics* [width=\textwidth] {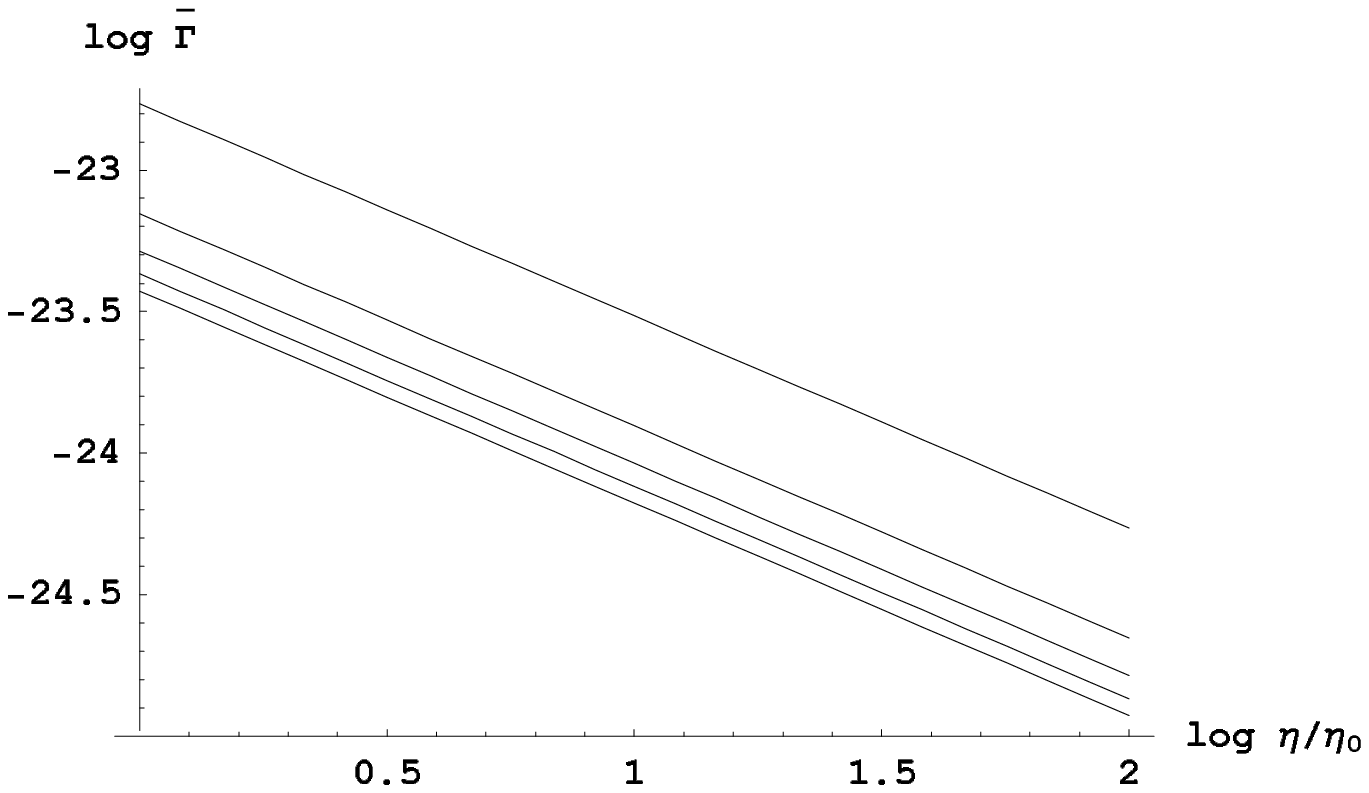}
\end{center}
\caption{Dependence of the reaction rate constant $\bar \Gamma$
 on solvent viscosity $\eta$  in the $log\ \bar \Gamma$ vs.
$log\ \left(\eta/\eta_0\right)$
coordinates given by formula (\ref{eq52}) for the case of
quartic (bistable) potential.
The values of $\left(A/D\right)_0$ from the down
line to the upper one respectively are: 1, 6, 11, 16, 21.}
\label{Fig.9}
\end{figure}

\end{document}